\documentclass[journal]{IEEEtran}

\usepackage{graphicx}  

\usepackage{amsmath}   
\begin{document}
%
\title{Magnetotransport in nonplanar SiGe/Si nanomembranes}
%
%
\author{G.~J.~Meyer,~N.~L.~Dias,~R.~H. Blick,
        and~I.~Knezevic
\thanks{This work was funded by the NSF
through the
University of Wisconsin MRSEC and award ECCS-0547415.}
\thanks{The authors are with the Department of Electrical and Computer Engineering, University of Wisconsin - Madison,
Madison, WI 53706, USA (email: knezevic@engr.wisc.edu).}
\thanks{Journal Reference: G. J. Meyer, N. L. Dias, R. H. Blick, and I. Knezevic, IEEE Trans. Nanotech. \textbf{6}, 446 (2007).}}
%
%
%
\markboth{IEEE Transactions on Nanotechnology,~Vol. 6, No.
4,~July~2007}{Meyer \MakeLowercase{\textit{et al.}}:
Magnetotransport in non-planar SiGe/Si nanomembranes}
%



\maketitle

\begin{abstract}
We investigate the relationship between electronics and geometry in
a nonplanar SiGe/Si resonant quantum cavity (RQC) subject to a
magnetic field. The transfer matrix technique originally due to
Usuki and coworkers [\textit{Phys. Rev. B}, vol. 52, pp. 8244-8255,
1995] has been modified to account for the nonzero local curvature
of the RQC. Our results demonstrate that low-temperature ballistic
magnetoconductance in nonplanar RQCs is highly sensitive to the
changes in curvature for a wide range of magnetic field strengths.
\end{abstract}


%
\IEEEpeerreviewmaketitle

\setcounter{page}{446}

\section{Introduction}
\PARstart{R}{ecent} interest in nonplanar two-dimensional (NP2D)
electron systems \cite{Leadbeater} has been driven by the emergence
of reliable techniques for fabrication of flexible heterostructures,
based on strained Si/SiGe \cite{Golod,Qin,Roberts} or GaAs
\cite{Lorke,Mendach}. Curvature has become another degree of freedom
available for manipulating electronic systems, which leads to novel
basic physics and suggests applications of NP2D
 systems in nanoelectromechanical systems (NEMS) as ultrasensitive scales and
sensors. The ability to design quantum devices using flexible
substrates increases with our understanding of the role played by
geometry in determining the electrical properties of these systems
\cite{Chaplik, Magarill, Chryssomalakos, Bertoni}. Therefore, there
is a need for versatile modeling tools capable of capturing the
physics in nonplanar structures in a variety of geometries, biasing
regimes, temperatures, and magnetic fields. To that end, we have
made modifications to a transfer-matrix technique \cite{Usuki,Ando},
widely and successfully implemented at low magnetic fields and in
planar 2D systems \cite{Sheng,Akis,Horie}, in order to expand its
functionality to include magnetic fields in non-planar geometries.

\begin{figure}
\centering
 \includegraphics[width=2.5in]{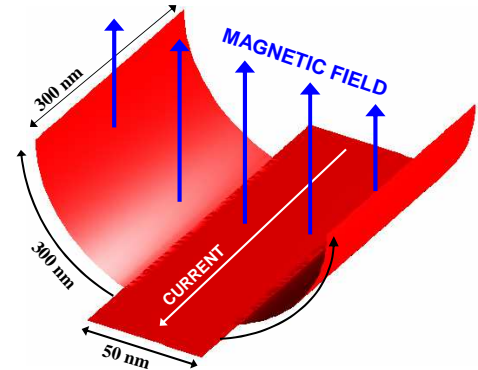}
\caption{\label{fig_RQC} A nonplanar resonant quantum cavity on a
SiGe/Si heterofilm, formed by selective etching of the sacrificial
layer under the cavity's wings. Strain from the lattice mismatch
provides the force that causes the wings to bow. Curvature is
measured terms of $\alpha=W / (\pi R$), where $W$ is the width of
the cavity and $R$ is the radius of curvature.}
\end{figure}

In this paper, we present a calculation of low-temperature ballistic
magnetoconductance in a nonplanar resonant quantum cavity (RQC)
under low bias, based on the modified transfer-matrix method. The
simulated 300$\times$300 nm$^2$ RQC (Fig. \ref{fig_RQC}) is formed
on a $\langle$100$\rangle$ SiGe/Si heterostructure upon selective
etching of the sacrificial oxide layer under the cavity's wings.
Narrow (50 nm) quantum-point contacts connect the cavity to the wide
leads (not depicted) at the source and drain ends;  the rigid leads
and the cavity spine are flat and tethered to the substrate. The
two-dimensional electron system is formed at the interface between
the undoped strained Si and the relaxed, phosphorous-doped SiGe
underneath (for example, epitaxially grown on SOI \cite{Roberts}).
The structure can be topped with another layer of n-type doped SiGe
to aid in the electronic confinement \cite{Roberts}; such SiGe/Si
modulation doped heterostructures have been shown to produce good
electronic confinement both before and after selective etching of
the sacrificial oxide layer. But unlike the strain-compensated
structure of Ref. \cite{Roberts} that remained flat upon
underetching, our structure releases the wings to curl up into a
partial cylinder, similar to what happens in the inverted Si/SiGe
structure described in Ref. \cite{Qin} that curled downward.
Curvature is measured in terms of a parameter $\alpha=W / (\pi R$),
where $W$ is the width of the cavity, and $R$ is the radius of
curvature.

In Sec. \ref{sec:method}, we introduce the modified transfer-matrix
method, applicable for low-bias transport in nonplanar structures
under magnetic field. Simulation results illustrating the effects of
curvature on transport are presented in Sec. \ref{sec:results}, and
concluding remarks are given in Sec. \ref{sec:conclusion}.

\section{The Modified Transfer-Matrix Method}\label{sec:method}
The transfer-matrix method due to Usuki \textit{et al.} \cite{Usuki}
is an iterative, stable routine used to solve the discretized
two-dimensional Schr\"{o}dinger equation. The entire numerical
domain is divided into slices perpendicular to the current flow. A
transfer matrix is constructed for each slice and, after iterating
slice-by-slice in the direction of the current flow, the
space-resolved electron density and the structure's transmission
coefficient are obtained. In the absence of a magnetic field,
\begin{eqnarray}
\label{eq:transfer} \left[ \begin{array}{c}
\psi_{n} \\
\psi_{n+1} \end{array} \right] &=& T_n \left[ \begin{array}{c}
\psi_{n-1}\\ \psi_{n} \end{array} \right], \\
T_n &=&\left[ \begin{array}{cc}
    \hat{0} & \hat{I} \\
    -\hat{H}^{-1}_{n,n+1}\hat{H}_{n,n-1} & \hat{H}^{-1}_{n,n+1}(E\hat{I}-\hat{H}_{n})
\end{array} \right], \nonumber \\
\label{eq:hamiltonian0} \hat{H}_{n,n+1}&=&\hat{H}_{n,n-1}
=\mbox{diag} [-E_0],\nonumber
\end{eqnarray}
where $\psi_n$ is the wavefunction of the $n$-th slice, $T_n$ is the
transfer matrix from slice $n-1$ to slice $n$, $\hat H_{n,n\pm 1}$
are the hopping Hamiltonian matrices that connect nearest-neighbor
slices, and $\hat H_n$ is the finite-difference representation of
the Hamiltonian of the $n$-th slice. All these Hamiltonians are
matrices of dimension $M$, where $M$ is the number of gridpoints in
the direction perpendicular to the current flow, while $\psi_n$ is
an $M$-column. Moreover, $E_0=\hbar^2/2m^*a^2$ is the
nearest-neighbor hopping energy, where $a$ is the mesh spacing
(assumed constant throughout the structure), and $m^*$ is the
electron effective mass. Since Si is biaxially strained in a
$\{$100$\}$ plane, among its 6 equivalent valleys the two whose
heavy mass is along the growth direction come down in energy with
respect to the other four (the so-called valley splitting into two
lower $\Delta_2$ and four higher $\Delta_4$ valleys \cite{Goswami}).
The electrons in $\Delta_2$ valleys respond to in-plane bias with
the light mass $m^{*}=0.19m_0$ ($m_0$ is the free-electron rest
mass), which is the effective mass we use in the calculation. The
subbands emerging from the $\Delta_2$ valleys upon confinement are
considered twofold degenerate (although the magnetic field can
introduce further splitting \cite{Goswami}). We will assume that
only the lowest (twofold degenerate) $\Delta_2$ subband is occupied.
The total electron sheet density in the leads, $2N$,  and the Fermi
level $E_F$ measured with respect to the subband bottom are
therefore connected by
\begin{equation}\label{eq:N}
N=\frac{m^*E_F}{\pi\hbar^2}.
\end{equation}

In order to calculate conductance though the structure, the
transmission coefficient for each propagating mode injected from the
source is calculated. At the source end of the device, there is a
wide lead (not depicted in Fig. 1), where many plane waves are
injected with unit amplitude into the device's active region (cavity
plus quantum-point contacts); only some make it all the way through
to the drain lead, with their amplitude significantly diminished.
Plane wave injection at the source end is achieved by solving Eq.
(\ref{eq:transfer}) for $n=0$ and $\psi_{1}=\psi_{0}e^{ika}$ (phase
shift between two adjacent slices will be $ka$ when a plane wave is
injected, where $k$ is the wave number), so
\begin{equation}
\label{eq:T0} e^{ika}\left[ \begin{array}{c}
\psi_{0} \\
\psi_{1} \end{array} \right] = T_0 \left[
\begin{array}{c}
\psi_{0} \\
\psi_{1} \end{array} \right].
\end{equation}
Since the elements of the transfer matrix $T_0$ are real [Eq.
(\ref{eq:transfer})] in the absence of magnetic field, the forward
and backward propagating modes occur in pairs: of the $2M$ modes
obtained, $M$ are forward and $M$ are backward modes; for each
direction, a small number of those $M$ modes are propagating and the
rest are evanescent, and the latter do not contribute to the
conductance. Only the forward propagating modes enter into the
conductance calculations, and thus must be selected from the
collection of $e^{ika}$ eigenvalues.  This is a straightforward
process in the absence of magnetic field: all propagating modes will
have real \emph{k} values, and can be identified as $|e^{ika}| = 1$.
Forward modes will have positive \emph{k} values, and thus
imag$(e^{ika}) > 0$. Note that the grid spacing \emph{a} must be
sufficiently small to ensure that all ${ka < \pi}$.

\subsection{Modifications for magnetic fields}

In the presence of a magnetic field, a gauge-invariant form of the
Schr\"{o}dinger equation with the magnetic field incorporated can be
obtained through the Peierls substitution \cite{Peierls}. This
substitution has the effect of producing the
$(\vec{\hat{p}}+e\vec{A})^{2}$ term in the full Hamiltonian, which
in itself is not suitable for implementation in transfer-matrix
approaches.

At zero magnetic field, one can obtain the matrix elements
${H}_{pq}$ of a tight-binding Hamiltonian by discretizing the
Schr\"{o}dinger equation. The effect of the magnetic field is
introduced by modifying each Hamiltonian matrix element $H_{pq}$ by
an appropriate Peierls phase factor, containing the integral of the
magnetic vector potential along a straight line connecting
meshpoints $p$ and $q$ \cite{Luttinger,Kohn,Hofstadter}:
\begin{equation}
\label{eq:Peierls} {H}_{pq} \rightarrow {H}_{pq}
\biggl(\frac{ie}{\hbar}\int_p^q{\vec{A}\cdot d\vec{\ell}}\biggr).
\end{equation}
In the Landau gauge $\vec A=(-By,0,0)$, the presence of the magnetic
field in the transfer matrix method is felt only in $\hat{H}_{n\pm
1}$:
\begin{eqnarray}\label{eq:hamiltonian1}
    \left(\hat{H}_{n,n+1}\right)_{jj}&=& -E_0\exp
\left(i 2\pi j\Phi/\Phi_0\right),\, j=1,\dots ,M,\nonumber\\
\hat{H}_{n,n-1}&=&\hat{H}^{\dagger}_{n,n+1},
\end{eqnarray}
where $\Phi=Ba^2$ is the flux through a meshcell, and $\Phi_0$ is
the magnetic flux quantum.

\begin{figure}
 \centering\includegraphics[width=2.5in]{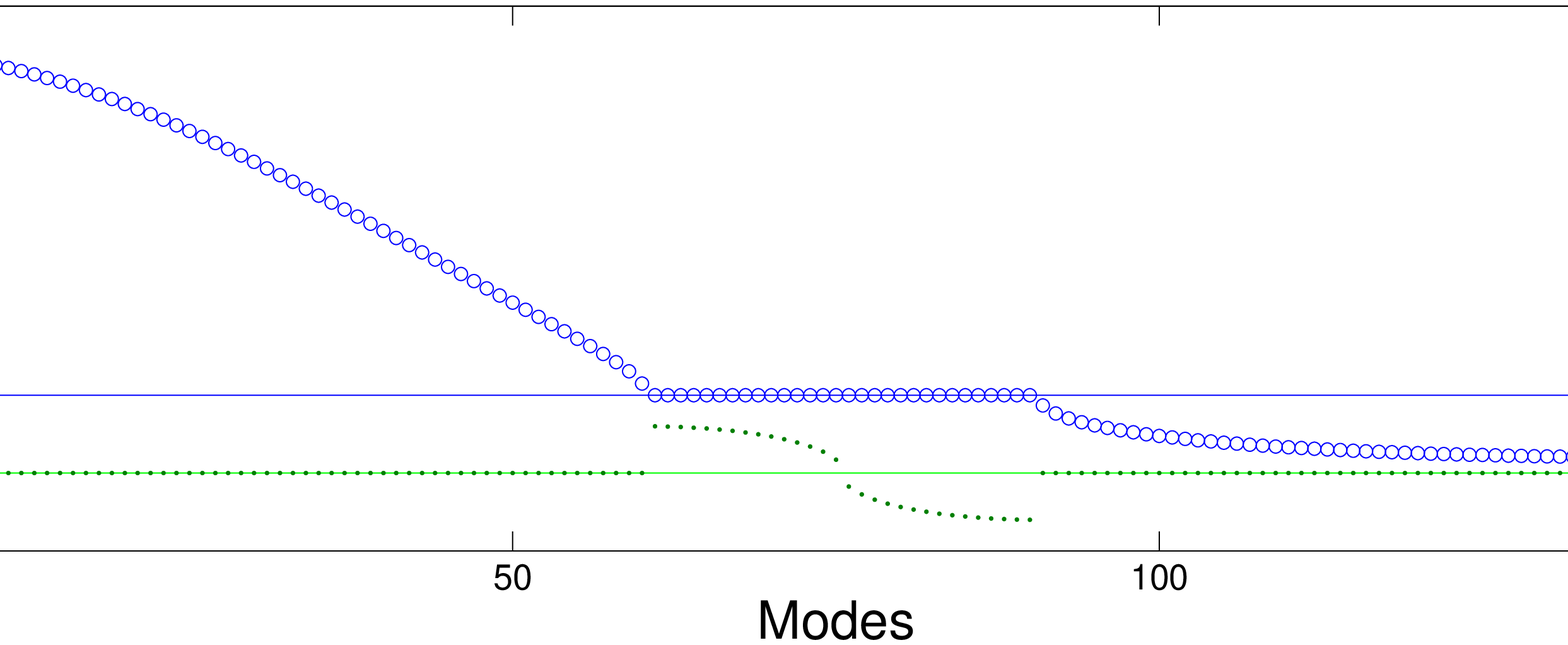}
 \centering\includegraphics[width=2.5in]{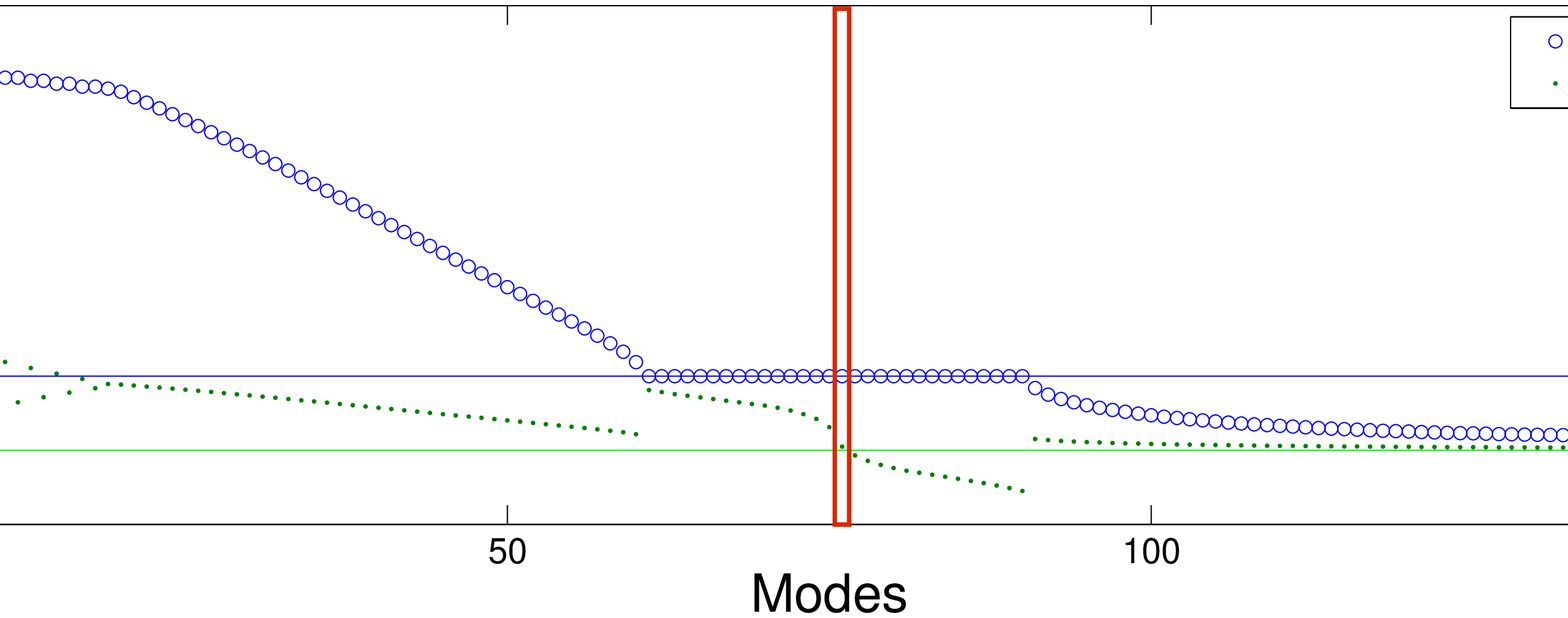}
 \centering\includegraphics[width=2.5in]{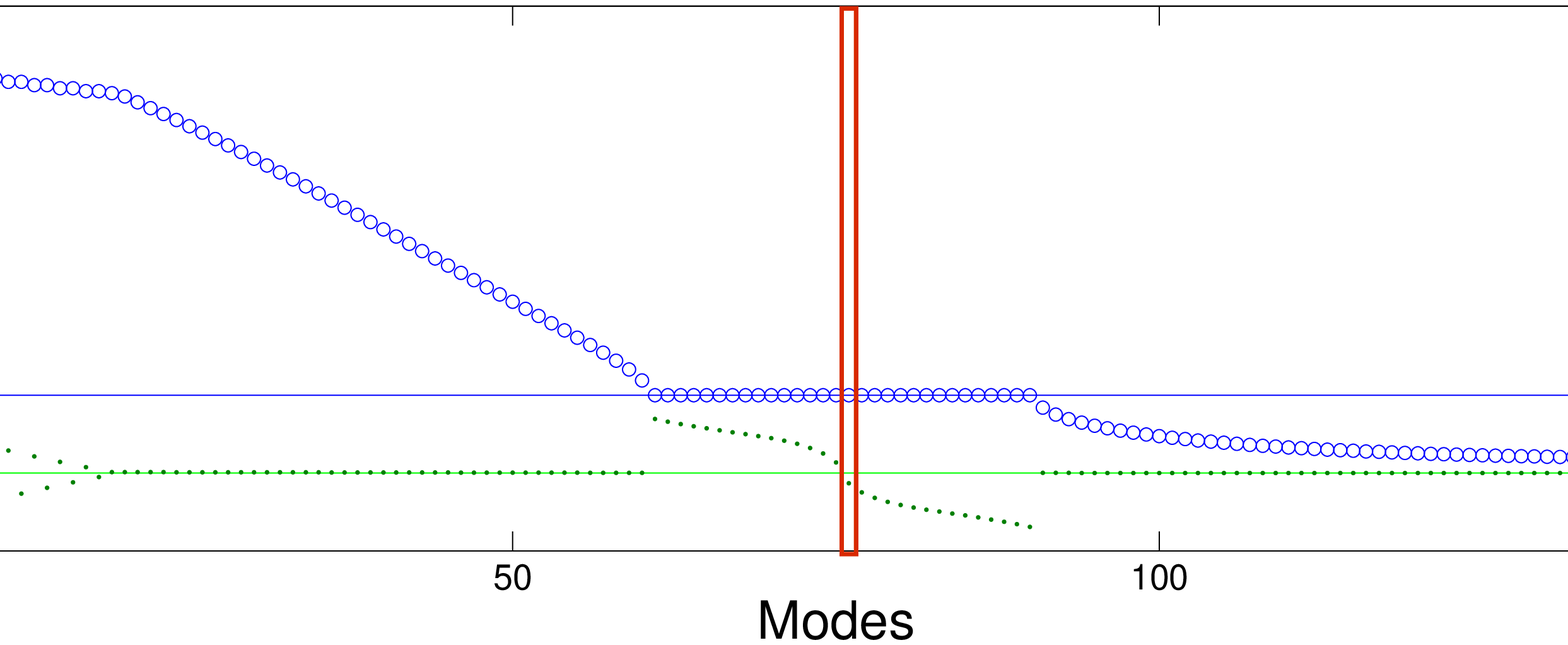}
\caption{ Absolute values and imaginary parts of all modes for
a 300 nm wire at $N=4\times 10^{11} \mathrm{cm}^{-2}$ and $B = 0$
(top graph) and $B = 0.4 T$ (middle and bottom graphs). The
Peierls phase skews the imaginary values of all eigenvalues and
erroneously reports an extra forward propagating mode (red box)
under this moderate magnetic field (middle graph). This skewing
effect is eliminated by taking $\int{\vec{A}\cdot d\vec{\ell}}$ from
the center of each stage (bottom graph) rather than the edge.}
\label{fig_modes}
\end{figure}

When the magnetic field is present, the matrix elements of $T_0$
become complex, and the forward and backward propagating modes are
no longer mutual complex conjugates. Since the essence of the
transfer-matrix technique is proper initialization of forward
propagating modes, the selection process becomes more complicated.
Namely, the eigenvalues $\mathrm{exp}{(ika)}$ the phase shifts
associated with the modes all get an extra positive phase shift
proportional to $\Phi/\Phi_0$, which is roughly an average of all
the Peierls' phases within one slice \cite{Ando}. Consequently, some
modes that are actually backward propagating may erroneously be
identified as forward propagating (see the middle panel of
Fig.\ref{fig_modes}). With increasing magnetic field, one can always
first try to decrease $a$, the meshsize, to minimize the phase
shift, but one quickly runs into problems with computer memory
(number of meshpoints in the structure and consequently the
dimensions of matrices dealt with become too large). Instead, we
have found that a simple shift of the coordinate origin -- placing
the origin at the center of the slice rather than the edge
(Fig.\ref{fig_modes})-- keeps this extra phase shift virtually
negligible up to very high magnetic fields. Certainly, this is not a
fundamental physical issue, but rather a particularity of the
transfer matrix technique.

\subsection{Modifications for nonplanar geometries} \label{sec:geom}

To investigate the effects of curvature combined with the presence
of a magnetic field, the transfer matrix implementation must be
further modified.

In the absence of a magnetic field, the motion of a particle
confined to a curved surface can be described in the surface's
natural coordinates, but with the addition of an extra attractive
geometric potential \cite{DaCosta}. The geometric potential $V_g$
depends on the mean and Gaussian curvatures of the surface at a
given point. For example, in the case of a cylindrical surface of
radius $R$, the geometric potential is a constant and is given by
$V_g= -\hbar^{2}/8m^{*}R^{2}$. Indeed, the transfer matrix method as
described in Sec. \ref{sec:method} is perfectly applicable to other
geometries, provided that one accounts for $V_g$, obeys the
structure's modified metric when constructing the kinetic energy
operator (details can be found in DaCosta's paper \cite{DaCosta}),
and discretizes the Schr\"{o}dinger equation in the surface's
natural coordinates.

When a magnetic field is added to the motion over a nonplanar
surface, the line integral $\int\vec{A}\cdot d\vec{\ell}$ that is
used in modifying the Hamiltonian matrix elements $H_{pq}$ must now
be taken \textit{over a geodesic between p and q} rather than along
a straight line. This implies that one first needs to calculate the
equations for the geodesics for the surface, and then integrate over
those to obtain the Peierls phase.

\section{Results}\label{sec:results}
The aforementioned modifications were made to the transfer matrix
method in order to observe the effects of curvature on the
electronic properties of the nonplanar RQC depicted in Fig. 1. Our
system is not fully cylindrical (it has cylindrical wings and a flat
spine) and is therefore more complicated than a cylinder due to the
varying geometric potential: $V_g$ is constant and attractive in the
wings, and zero in the spine. (An interested reader can find the
analytical solution to magnetotransport on a cylinder with very
small curvature in Ref. \cite{Lorke}.)

In Figs. 3-6, conductance variations in the structure (in units of
$G_0=2e^2/h$, the conductance quantum)  are presented. All the
results are obtained in the low-bias regime as well as at low
temperatures, so the electron transport occurs at the Fermi level,
specified by $N$, the electron sheet density (per $\Delta_2$ valley)
in the source and drain leads [Eq. (\ref{eq:N})]. Conductance
calculated is also per $\Delta_2$ valley. Conductance plots are
accompanied by color plots of the magnitude squared of the
wavefunction at the Fermi energy, with low/moderate/high values
corresponding to the usual blue/yellow/red color scheme. (This is
the same wavefunction that determines the transmission coefficient.
It involves many plane waves being injected with unit amplitude from
the source lead into the structure's active region, as described in
Section II. )

\subsection{Zero Magnetic Field} 
\begin{figure}
 \centering\includegraphics[width=3in]{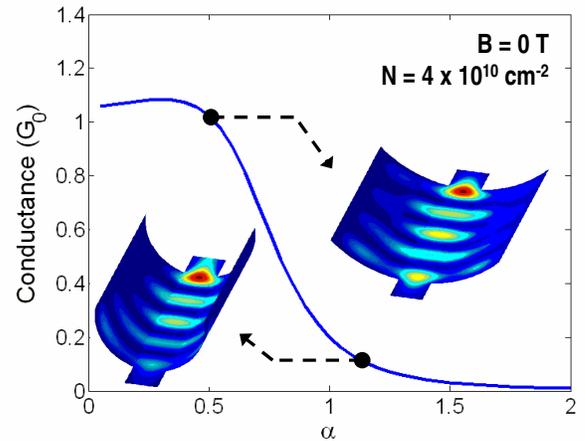}
\caption{Conductance variation with curvature at $B=0$. At
particular electron densities in the leads, the conductance is
highly dependent on the curvature of the cavity even in the absence
of a magnetic field. This phenomenon can only be attributed to the
geometric potential. Insets: colorized plots of the magnitude
squared of the wavefunction at the Fermi energy, for the specific
values of $\alpha$ shown.} \label{fig_noflux}
\end{figure}

The geometric potential defined in \ref{sec:geom}
is quite weak at the dimensions of our RQC, and its effect often
goes undetected.  At certain sheet electron densities $N$, however,
it alone is responsible for significant variation in conductance.
Hence, conductance can be modulated by changing the curvature even
if in the absence of a magnetic field, as shown in Fig.
\ref{fig_noflux}. At lower curvatures, most of the plane waves
incident from the source lead (top) are transmitted to the drain
[note the high probability density in the drain-end quantum-point
contact (right-hand-side inset)]. As the curvature increases, the
coupling between the source and drain decreases, leading to a
decrease in conductance.

\subsection{Low Magnetic Field}
\begin{figure}
 \centering\includegraphics[width=3in]{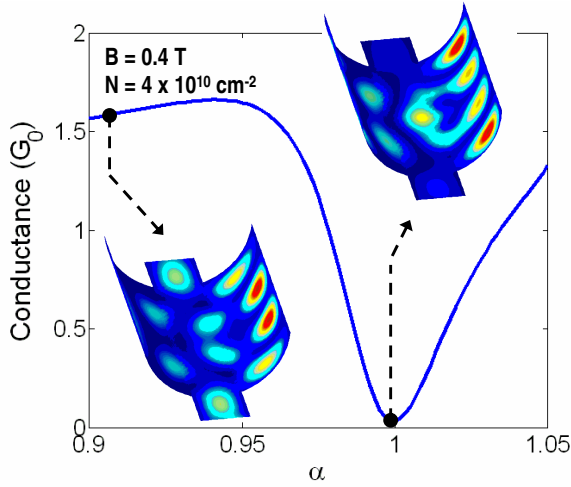}
\caption{Conductance variation with curvature at a low magnetic
field ($B=0.4$T). Insets: colorized plots of the magnitude squared
of the wavefunction at the Fermi energy, for the specific values of
$\alpha$ shown. Patterns observed in the probability density are
altered by changes in curvature, significantly affecting the
conductance.} \label{fig_lowflux}
\end{figure}

At low magnetic fields, the resonance patterns which form in the
cavity on account of the magnetic field  \cite{Akis,Horie} can be
altered by cavity curvature. These patterns determine the
transmission through the cavity (see Fig. \ref{fig_lowflux}), and
hence alter the conductance as well.

\subsection{High Magnetic Field}

\begin{figure}
 \centering\includegraphics[width=2.5in]{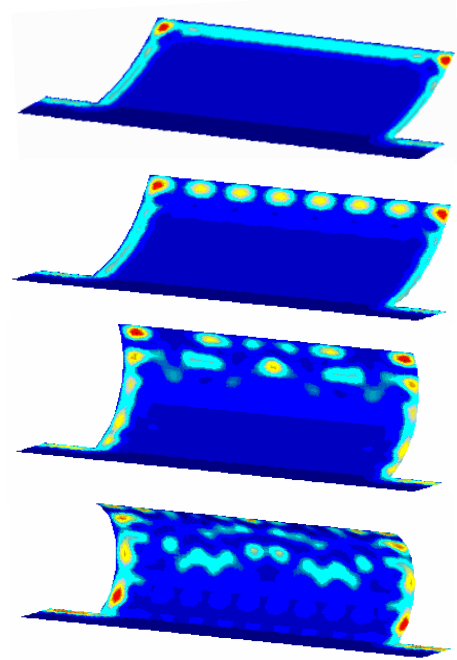}
\caption{Magnitude squared of the wavefunction at the Fermi energy
in the nonplanar RQC as a function of curvature at a high magnetic
field ($B = 12.5$T). $N=4\times10^{11} cm^{-2}$. At low curvature
($\alpha=0.4$, top panel) the edge state is clearly visible.  As
curvature increases ($\alpha=0.5$, second panel), the edge state
migrates away from the edge of the cavity and broadens. As the
cavity takes on a distinctly cylindrical shape ($\alpha$=1.5, third
panel and $1.7$, bottom panel), it exhibits a mid-cavity state,
created by opposing magnetic fluxes above and below it. Near wing of
the cavity, where the probability density is effectively zero, has
been removed so it would not obscure the images.}
\label{fig_edgeStates}
\end{figure}

\begin{figure}
 \centering\includegraphics[width=3in]{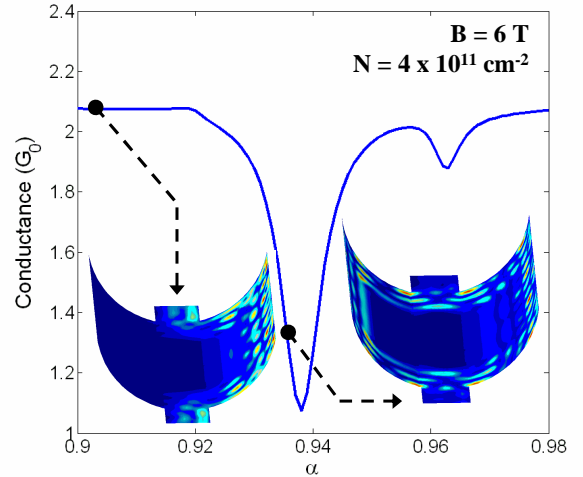}
\caption{\label{fig_highflux}Conductance variation with curvature at
a high magnetic field ($B=6$T). Insets: colorized plots of the
magnitude squared of the wavefunction at the Fermi energy, for the
specific values of $\alpha$ shown. At certain magnetic fields, such
as the one presented here, the cavity allows edge states which are
prohibited in the narrow quantum-point contact. The shape of the
cavity determines how these edge states couple to the modes in the
contact; therefore, conductance again varies with curvature.}
\end{figure}

At high magnetic fields, the results of our calculations confirm
what is known about the physics of high-field magnetotransport in
planar structures, such as the formation edge states and
depopulation of Landau levels. In addition, novel effects of
curvature on current conduction in the strong quantum Hall regime
are also observed (see Fig.\ref{fig_edgeStates}). At high fields,
the edge states (top panel of Fig. \ref{fig_edgeStates}) become
wider and drift away from the edge of the cavity as the cavity rolls
up (its curvature increases). As the cavity takes on a distinctly
cylindrical shape, a mid-cavity state is formed as a result of
opposing fluxes through the top and bottom portions of the wing.

At certain magnetic fields (Fig. \ref{fig_highflux}), even minuscule
changes in the curvature cause significant changes in the
conductance. The conductance will remain sensitive to the changes in
curvature in the high-field regime up to extremely high magnetic
fields ($B
> 8T$ in our particular example), after which the edge states
become increasingly difficult to disturb by changes in curvature.

\section{Conclusion}\label{sec:conclusion}
We have developed a modified transfer-matrix method that enables
low-bias, low-temperature transport calculations for nonplanar
two-dimensional electron systems in the presence of magnetic fields.
We have shown that the interplay between geometry and electronic
properties of flexible SiGe/Si heterostructures (such as the
simulated cylindrical resonant quantum cavity) is substantial. In
particular, our results demonstrate that conductance through a
cylindrical RQC at a given carrier density in the leads is highly
sensitive to the changes in curvature for a wide range of magnetic
field strengths, which suggests possible applications of nonplanar
structures as nanoelectromechanical systems (NEMS) and sensors
(e.g., a dilute chemical presence that affects the shape of the film
could be detected electronically), at least at very low
temperatures. At higher temperatures, scattering due to phonons must
not be disregarded, and energies within a few $k_BT$ around the
Fermi level will contribute to the conductance, which may require
including several $\Delta_2$ subbands as well as some low-lying
$\Delta_4$ subbands. Consequently, smearing of the resonant features
in $G$ vs $\alpha$ (Figs. 3, 4, and 6) ought to be expected.
Exploration of non-cryogenic transport in nonplanar RQCs will be the
topic of a subsequent publication.

\end{document}